\documentclass[prd,aps,floats,tabularx,preprintnumbers,twocolumn]{revtex4}
\usepackage{graphicx, epsfig}

\newcommand{\be}{\begin{equation}}
\newcommand{\ee}{\end{equation}}
\newcommand{\bea}{\begin{eqnarray}}
\newcommand{\eea}{\end{eqnarray}}
\newcommand{\vis}{{\rm vis}}

\newcommand{\PRL}{{\it Phys. Rev. Lett.\,}}

\def\fun#1#2{\lower3.6pt\vbox{\baselineskip0pt\lineskip.9pt
        \ialign{$\mathsurround=0pt#1\hfill##\hfil$\crcr#2\crcr\sim\crcr}}}

\newcommand\lsim{\mathrel{\rlap{\lower4pt\hbox{\hskip1pt$\sim$}}
    \raise1pt\hbox{$<$}}}
\newcommand\gsim{\mathrel{\rlap{\lower4pt\hbox{\hskip1pt$\sim$}}
    \raise1pt\hbox{$>$}}}

\def\dslash{\not{\hbox{\kern-2pt $\partial$}}}
\def\Dslash{\not{\hbox{\kern-4pt $D$}}}
\def\Oslash{\not{\hbox{\kern-4pt $O$}}}
\def\Qslash{\not{\hbox{\kern-4pt $Q$}}}
\def\pslash{\not{\hbox{\kern-2.3pt $p$}}}
\def\kslash{\not{\hbox{\kern-2.3pt $k$}}}
\def\qslash{\not{\hbox{\kern-2.3pt $q$}}}

 \newtoks\slashfraction
 \slashfraction={.13}
 \def\slash#1{\setbox0\hbox{$ #1 $}
 \setbox0\hbox to \the\slashfraction\wd0{\hss \box0}/\box0 }

\def\ee{\end{equation}}
\def\be{\begin{equation}}

\newcommand{\cvis}{c_{\text{vis}}^2}

\begin{document}

\setlength{\unitlength}{1mm}
\title{Anisotropies in the Cosmic Neutrino Background after WMAP 5-year Data}

\author{Francesco De Bernardis$^1$, Luca Pagano$^1$, 
Paolo Serra$^2$, Alessandro Melchiorri$^1$, Asantha Cooray$^2$}

\affiliation{$^1$Physics Department and sezione INFN, University of
  Rome ``La Sapienza'', Ple Aldo Moro 2, 00185 Rome, Italy\\
$^2$Dept. of Physics \& Astronomy University of California 
Irvine, Irvine, CA 92697-4575}

\date{\today}%


\begin{abstract}

We search for the presence of cosmological neutrino background (CNB) anisotropies in
recent WMAP 5-year data using their signature imprinted on modifications to 
cosmic microwave background (CMB) anisotropy power spectrum.
By parametrizing the neutrino background anisotropies with the speed viscosity 
parameter $c_\vis$, we find that the WMAP 5-year data alone provide
only a weak indication for CNB anisotropies with $c_\vis^2 > 0.06$ at the
$95 \%$ confidence level. When we combine CMB anisotropy data with measurements of
galaxy clustering, SN-Ia Hubble diagram, and other
cosmological information, the detection increases to 
$c_\vis^2 > 0.16$ at the same $95 \%$ confidence level. Future data from Planck, 
combined with a weak lensing survey such as the one expected with DUNE from space, 
will be able  to measure the CNB anisotropy parameter at about $10\%$ accuracy.
We discuss the degeneracy between neutrino background ansiotropies and other cosmological parameters
such as the number of effective neutrinos species and
the dark energy equation of state.
\end{abstract}
\bigskip

\maketitle

\section{Introduction}

The recent results on the cosmic microwave background (CMB) anisotropies 
from the the five-year data of the WMAP mission
have once again confirmed the basic predictions of the standard
cosmological model \cite{Komatsu:2008hk}. In addition to
an improved determination of several key cosmological parameters such as the scalar spectral index 
and the optical depth to reionization with a better control of systematics, the new WMAP 5-year data
have provided, probably for the first time, a clear indication for the presence of ansiotropies
in the cosmic neutrino background (CNB) at more than $95 \%$ confidence level.

The CNB is a clear prediction of the standard cosmological model, though a direct detection of
the cosmic neutrinos remain extremely challenging. These neutrinos decouples from the primeval
plasma at temperature $T\sim 1$ MeV before electron-positron annihilation. A relic neutrino background is
expected today at a temperature of $T_{\nu}=(4/11)^{1/3}T_{\gamma}$ (where $T_{\gamma}=2.728 K$
is the CMB blackbody temperature)  with an energy density of
$\rho_{\nu} \sim 0.58 N_{\rm eff} T_{\nu}^4$.
In the standard cosmological model $N_{\rm eff}=3.045$ and any hint
for $N_{\rm eff}>0$ can be therefore considered as an indication for the CNB.

Recent combined analyses of CMB data with other cosmological
observables have already provided striking evidence for the  presence of this neutrino background. 
A combination of WMAP three-year data with measurements of Hubble expansion rate, for example,
provided the constraint that $N_{\rm eff}=3.7\pm1.1$ at $95 \%$ confidence level \cite{debeage}, with
other analyses finding consistent results (see e.g. \cite{neutrinos}).
The recent WMAP 5-year data analysis has reported an estimate of $N_{\rm eff}$, but only inluding
CMB data in the analysis \cite{Komatsu:2008hk}. Their results is then the most conservative, though it is now clear that neutrinos
are certainly an important energy component of our universe.

In this paper we reanalyze, in light of the new WMAP 5-year data release and their results,
some further properties of the neutrino background, namely
the presence of anisotropies in the background and not just the effective number of neutrino species. 
Although inflationary anisotropies in the CNB at the level of $\sim 10^{-5}$ are expected
in the standard scenario, a direct detection of this anisotropy is significantly more challenging
than a simple detection of the relic CNB today. The expected anisotropies in the CNB, however, affect the
CMB anisotropy angular power spectrum at level of $\sim 20 \%$
through the gravitational feedback of the neutrino free-streaming damping
and anisotropic stress contributions \cite{Huetal95}. This allows an
indirect detection of the neutrino background ansiotropy through its signature in the CMB.

A way to parameterize the anisotropies in the
CNB has been introduced in \cite{Hu98} with the ``viscosity
parameter'' $c_{\vis}^2$, which controls the relationship between
velocity/metric shear and anisotropic stress in the CNB. 
In the standard scenario $c_{\vis}^2=1/3$,
 anisotropies are present in the CNB and approximate the
radiative viscosity of neutrinos. The case
$c_{\vis}^2=0$, on the contrary, cuts the Boltzmann hierarchy of
CNB perturbations at the quadrupole, forcing a perfect fluid
solution with no CNB anisotropies but only density and velocity
(pressure) perturbations.
Observationally determining $c_{\vis}^2>0$ would
therefore provide a strong indication for the existence of CNB
anisotropies, as argued in \cite{Huetal98}. This possibility is studied in several 
papers that made use of early CMB data \cite{cvis}  with varying levels of
indirect detection of the CNB anisotropy.

In this paper we reanalyze the constraints on neutrino anisotropies
in light of the new WMAP 5-year data. We combine CMB data with existing large-scale structure cosmological
information to get an overall estimate on the viscosity parameter and to constraint it better, for the first time,
at a confidence level greater than 95\%.  Moreover, we also study how future experiments as the Planck satellite
CMB mission, possibly combined by future weak lensing data, will be able to constrain this parameter. 
Our paper is organized as follows: in the next Section, we discuss the CMB and large-scale structure data analysis
and presenr our results based on existing data in Section~III. We also forecast the expected errors in Section~IV and conclude
with a summary of our results in Section~V.

\section{Analysis}

The method we adopt for this analysis is based on the publicly available Markov Chain Monte Carlo
package \texttt{cosmomc} \cite{Lewis:2002ah} with a convergence
diagnostics based on the Gelman and Rubin statistic.
We sample the following eight-dimensional set of cosmological
parameters, adopting flat priors on them: the physical baryon and Cold Dark Matter,
$\omega_b=\Omega_bh^2$ and $\omega_c=\Omega_ch^2$,
 the ratio of the sound horizon to the angular diameter
distance at decoupling, $\theta_s$, the scalar spectral index
$n_S$, the overall normalization of the spectrum $A$ at $k=0.05$
Mpc$^{-1}$, the optical depth to reionization, $\tau$, the number
of massless neutrinos $N_{\rm eff}$ and, finally, the viscosity
parameter $c_\vis$ introduced above with the prior $c_{\rm vis}^2 \le
1/3$. Simultaneously  we also use a cosmic age top-hat prior as 10 Gyr$ \le t_0 \le$ 20 Gyr.

Furthermore, we consider purely adiabatic initial conditions, we impose flatness and we
treat the dark energy component as a cosmological constant.
We include the five-year WMAP data \cite{Komatsu:2008hk} (temperature
and polarization) with the routine for computing the likelihood
supplied by the WMAP team. Together with the WMAP data we also
consider the small-scale CMB measurements of ACBAR \cite{Reichardt:2008ay},
CBI \cite{2004ApJ...609..498R}, and BOOMERANG-2K
\cite{2005astro.ph..7503M}.  In addition to the CMB data, we
include the Supernovae Legacy Survey data \cite{2006A&A...447...31A}, 
the real--space power spectrum of 
galaxies from the Sloan galaxy redshift survey (SDSS) \cite{2004ApJ...606..702T}, and 
 2dF galaxy power spectrum \cite{2005MNRAS.362..505C}, and, in a separate analysis, constraints from the real--space
 power spectrum of red galaxies from the Sloan galaxy
 redshift survey (SDSSlrg) \cite{Tegmark:2006az}.

\begin{figure}[t]
\begin{center}
\includegraphics[width=\linewidth]{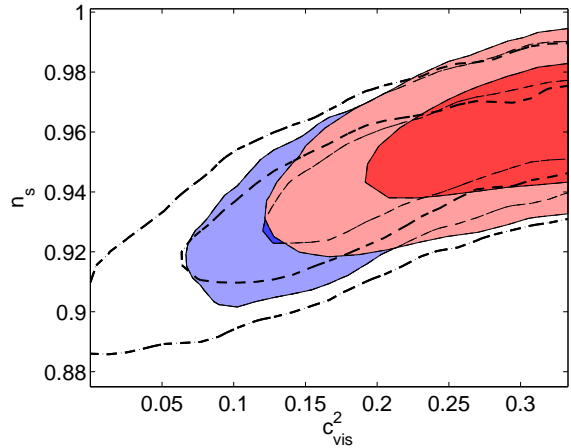}
\caption{Joint 2-dimensional posterior probability contour plots
in the $c_{\vis}^2-n_s$ plane, showing the $68\%$ and $95\%$ confidence level
 contours from the WMAP 5-year data alone (empty contours), ALL+SDSSlrg 
dataset (blue contours) and ALL+SDSS dataset (red contour). 
Including the full Sloan dataset is necessary to improve
the limits placed on $c_\vis$} \label{cvisVSns}
\end{center}
\end{figure}

Our analysis considers 2 different scenarios: firstly, we test for
anisotropies in a relativistic background comprising $N_{\rm eff}=3$
massless neutrino species. Secondly, we allow the number of
massless neutrinos to vary. In both cases, we allow for variations in
the 
viscosity parameter $c_\vis$.

We consider here 3 separate datasets: the first with WMAP 5-year only (WMAP5), the second 
with all existing CMB data (ALL) combined with data from Sloan
red galaxy power spectrum (ALL+SDSSlrg), and the third with
all CMB information combined with the SDSS power spectrum (ALL+SDSS). 
In the last two cases we have also imposed the Guassian Hubble Space Telescope 
prior\cite{Freedman:2000cf}: $h = 0.72\pm0.08$ and also a weak 
big-bang nucleosynthesis prior\cite{Burles:2000zk}: 
$\omega_bh^2 = 0.022 \pm 0.002 \quad (1 \sigma)$. In the computation
of the CMB spectra we have allowed for the lensing modification \cite{lensing}. 
As above for the other parameters we have adopted flat priors.

\section{Results}

\subsection{Standard background of $N_{\rm eff}=3$ massless
neutrinos.}

In Figure~\ref{cvisVSns} we plot constraints obtained from our
analysis in the $c_{\vis}^2-n_s$ plane in the case of $N_{\rm eff}=3$
massless neutrinos. The WMAP 5-year data alone is able to put only a weak 
constrain on $c_\vis$, due to the degeneracy with 
the spectral index highlighted in \cite{melktrot}, $c_{\vis}^2 > 0.06$ at $95\%$ confidence level.
Including the full cosmological dataset and the red galaxy spectrum (ALL+SDSSlrg)
the degeneracy is partially broken enough to obtain $c_{\vis}^2 > 0.11$  (95\% confidence level). Including the full Sloan + 2dF dataset (ALL+SDSS)
we obtain a stronger constrain with $c_{\vis}^2 > 0.16$ at the same 95\% confidence level.
The difference in log--likelihood between the best fit for 
the standard case and the
case with no anisotropies ($\cvis=0$) is $\Delta \chi^2 = 15.2$.
Therefore we can conclude that the case where the NB does not have
anisotropies above the first moment is quite clearly disfavored.

\subsection{General background of massless neutrinos.}

\begin{figure}[t]
\begin{center}
\includegraphics[width=\linewidth]{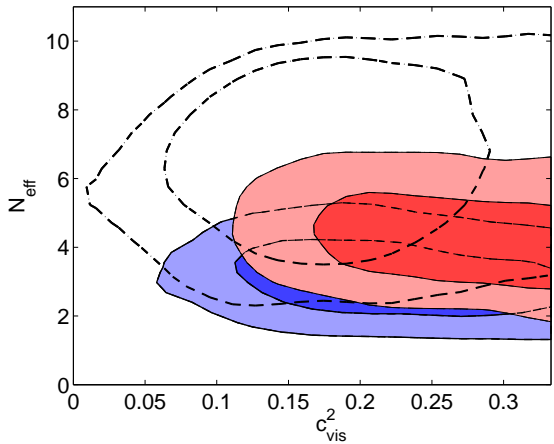}
\caption{Joint 2-dimensional posterior probability contour plots
in the $c_{\vis}^2-N_{\rm eff}$ plane, showing the $68\%$
and $95\%$  contours from the WMAP5 data alone (empty contours), ALL+SDSSlrg 
dataset (blue contours) and ALL+SDSS dataset (red contour). \label{cvisVsNnu}}
\end{center}
\end{figure}

We now relax the assumption of a standard background of
relativistic neutrinos by treating $N_{\rm eff}$ as a free parameter and
thus check the ability of the data to put simultaneous constraints
both on the background number of neutrinos and on the
presence of anisotropies in it. In Figure~\ref{cvisVsNnu} we show the
2--dimensional constraints in the plane $c_{\vis}^2-N_{\rm eff}$. 

\begin{table}[htb]\footnotesize
\begin{center}
\begin{tabular}{l|c|cc}
 Dataset& $c_{\vis}^2 \quad $ $N_{\rm eff}=3$  & $N_{\rm eff}$ & $c_{\vis}^2$  \\
\hline
WMAP5 & $\ge 0.06$ &$6.4^{+2.8}_{-3.2}$ & $\ge 0.07$ \\
ALL+SDSSlrg& $\ge 0.11 $ &$3.2^{+1.7}_{-1.5}$ & $\ge 0.11$  \\
ALL+SDSS& $\ge 0.16 $ &$4.4^{+2.1}_{-1.8}$ & $\ge 0.15$ \\
\hline
\end{tabular}
\caption{The $95\%$ confidence level limits on $N_{\rm eff}$ and lower limit at $95\%$ 
c.l. on $c_{\vis}^2$ for WMAP5 data alone, ALL+SDSSlrg and ALL+SDSS (see text for details).}
\label{resultsNnuvar}
\end{center}
\end{table}

As we can see, the WMAP data alone can
provide a weak indication for for a background of relativistic
neutrinos, giving $3.2 < N_{\nu} < 9.2$ at $95 \%$ c.l., and it is able to put also 
a weak constraint on $c_\vis$, giving $c_{\vis}^2 > 0.07$ at $95 \%$ c.l.. 
Considering the ALL+SDSSlrg dataset we obtain $1.7 < N_{\nu} < 4.9$ at $95 \%$ c.l.
and $c_{\vis}^2 > 0.11$ at $95 \%$ c.l.. 
Including the full Sloan dataset increases the constraining
on $c_{\vis}^2$:

 \bea
 \cvis > 0.15 & \quad (95\% \text{ c.l., 1--tail}) \\
 2.6 < N_{\nu} < 6.5 & \quad (95\% \text{ c.l., 2--tails}).
 \eea

The difference in log--likelihood between the best fit for 
the standard case and the
case with no anisotropies ($\cvis=0$) is $\Delta \chi^2 = 19.4$.

\section{Forecast}

In this section we investigate how future experiment could improve the
constraints on the viscosity parameter 
$c^{2}_{\vis}$, the number of relativistic species $N_{\rm eff}$ and the
scalar spectral 
index $n_s$ together with other cosmological parameters. 
Several future experiments will focus on the observation of
weak gravitational lensing  with large-area galaxy surveys.
While weak lensing surveys can improve our understanding of neutrino
physics \cite{Cooray2}, 
weak lensing observations help in this analysis
by providing a way to break certain degeneracies by improving the determination of
parameters such as $n_s$. CMB experiments, however, are  more
sensitive to $c^{2}_{\rm vis}$, though in CMB alone, this parameter is degenerate with other quantities. 
It is therefore interesting to study how the combination  of a planned CMB experiment, mainly Planck, with weak lensing
surveys can  improve the constraints on the CNB anisotropy
parameters. 

\subsection{Fisher matrix}

To explore this
we perform a Fisher matrix analysis \cite{Jungman:1995bz} where
the Fisher matrix is
\begin{equation}\label{fisher}
F_{\alpha\beta}\equiv\langle-\frac{\partial^{2}\ln L}{\partial p_{\alpha}\partial p_{\beta}}\rangle,
\end{equation} 
where $L$ is the likelihood function of the set of parameters
$p_{i}$.  Here we adopt a cosmological  model with ten parameters. We combine the Fisher matrix of Planck
with the  Fisher matrix of three different weak lensing surveys to study the
dependence of the constraints on the survey parameters since weak lensing surveys are still under
development. According to the Cramer-Rao inequality
for unbiased estimators this is the best 
statistical error that one can obtain for the generic parameter $p_{\alpha}$:
\begin{equation}\label{sigma}
\sigma_{p_{\alpha}}\geq\sqrt{(F^{-1})_ {\alpha\alpha}}.
\end{equation}

For a CMB experiment the Fisher matrix is given by:
\begin{equation}
     F^{CMB}_{\alpha\beta} = \sum_{l=2}^{l_{\rm max}} \sum_{i,j}
     \frac{\partial C_l^{i}}{\partial p_\alpha}
     ({\rm Cov}_l)_{i j}^{-1}
     \frac{\partial C_l^{j}}{\partial p_\beta},
\label{fishercmb}
\end{equation}
where the $C_l^{\alpha\beta}$ are the well known power spectra for the
temperature (TT), 
temperature-polarization (TE), E-mode polarization (EE), and B-mode
polarization ($i$ and $j$ run over TT, 
EE,TE, BB) and ${\rm Cov}_l$ is the spectra covariance matrix. We use
the  experimental configuration of the 
Planck satellite to compute the Fisher matrix for CMB
anisotropies. The specifications for this experiment are listed in 
Table \ref{Planck}. We use $l_{\rm max}=1500$ to calculate the sum in (\ref{fishercmb}).

\begin{table}[htb]\footnotesize
\begin{center}
\begin{tabular}{rcccc}
& Chan. & FWHM & $\Delta T/T$ & $\Delta P/T$  \\
\hline
$f_{\rm sky}=0.65$
& 100 & $9.5'$ & 2.5 & 4.0 \\
& 143 & $7.1'$ & 2.2 & 4.2 \\
& 217 & $5.0'$ & 4.8 & 9.8 \\
\hline
\end{tabular}
\caption{Experimental specifications for the Planck satellite. Channel frequency is given in GHz, FWHM in arcminutes, and noise in $10^{-6}$.}
\label{Planck}
\end{center}
\end{table}

The future weak lensing surveys allow the possibility to
''tomographycally'' reconstruct the distribution 
of foreground matter responsible for distortions in the images of distant
background galaxies \cite{Hannestad:2006as}. One can
therefore divide the survey into various redshift 
bins and reconstruct the convergence weak lensing power spectrum
$P_{ij}(\ell)$ at multipole $\ell$; 
the subscripts $i$ and $j$ are referred to the redshift bins. The
convergence power spectrum contains the non linear 
matter power spectrum $P_{nl}$ at redshift $z$, obtained correcting
the linear power spectrum $P(k,z)$. 

With these assumptions the Fisher matrix for weak lensing is given by \cite{Amendola:2007rr}:
\begin{equation}\label{fisherwl}
    F_{\alpha\beta}=f_{sky}\sum_{\ell}\frac{(2\ell+1)\Delta\ell}{2}\frac{\partial
    P_{ij}}{\partial p_{\alpha}}C_{jk}^{-1}\frac{\partial
    P_{km}}{\partial p_{\beta}}C_{mi}^{-1},
\end{equation}
where $\Delta\ell$ is the step used for $\ell$ and:
\begin{equation}\label{c}
    C_{jk}=P_{jk}+\delta_{jk}\langle\gamma^2_{int}\rangle n_j^{-1} \, ,
\end{equation}
is the lensing covariance. Here, we ignore non-Gaussian contribution to the covariance \cite{CooHU} as we are interested in
obtaining an estimate on the ability of these experiments to measure CNB parameters.

In the last expression $\gamma_{int}$ is the rms intrinsic shear (and
we assume $\langle\gamma^2_{int}\rangle^{1/2}=0.22$ consistent with high fidelity space-based imaging data and future ground-based data)
and $n_j$ is the number of galaxies per steradian belonging to $j$th bin:
\begin{equation}\label{number}
    n_j=3600d\left(\frac{180}{\pi}\right)^2\hat{f}_{j} \, ,
\end{equation}
where $d$ is the number of galaxies per square arcminute and $\hat{f}_j$ is the fraction of sources belonging to the
$j$th bin. In our analysis we use $\ell$ in the range
$10\leq\ell\leq2000$ , so that we are well within the linear to mildly non-linear regime of
 fluctuations and not to bias our results with uncertainties in
 predictions related to non-linear regime at $\ell>2000$.. 
We considered three different future weak
lensing surveys, PanSTARRS \cite{pan}, DES \cite{des} 
and DUNE \cite{dune} experiments, with different sky coverage
($f_{sky}$), median redshift $z_{0}$ and number 
of sources per square arcminute $d$. We assume a radial distribution function of galaxies given by $D(z)=z^2\exp[-(z/z_0)^{1.5}]$
where $z_0$ depends on the survey considered. The experimental
characteristics of these survey are listed in table \ref{survey}.

Here, we account for uncertainties in the measured photometric
redshifts of the galaxies,  according to the approach of \cite{Ma:2005rc}: if $p(z_{ph}|z)$ is
the probability that a galaxy with (real) redshift $z$ is observed
at photometric redshift $z_{ph}$ then the distribution of galaxies
in the $i$th bin is given by $D_i(z)=\int_{z_{ph,i}}dz_{ph}D(z)p(z_{ph}|z)$.
We choose the probability $p(z_{ph}|z)$ to be Gaussian: 
$p(z_{ph}|z)=\frac{1}{\sqrt{2\pi}\sigma_z}exp\left[-\frac{(z_{ph}-z)^2}{2\sigma_z^2}\right]$ 
with $\sigma_z$ depending on the survey.

We then combine the Fisher matrix of Planck with the Fisher matrix of
different weak lensing surveys to show how these 
surveys can improve the constraints coming from CMB. Since CMB
 anisotropies and weak lensing of distant 
galaxies have origin in two distant epochs of the history of the
 Universe we can consider them 
to be independent. Then, the total Fisher matrix is simply the sum of the Fisher matrix of CMB and that of weak lensing:
\begin{equation}\label{sum}
F^{TOT}_{\alpha\beta}=F^{CMB}_{\alpha\beta}+F^{WL}_{\alpha\beta} \, .
\end{equation}

\begin{table}[htb]\footnotesize
\begin{center}
\begin{tabular}{ccccc}
experiment & $f_{sky}$ & $z_0$  & $d$ & $\sigma_{z}$\\
\hline
\hline
DES & $0.13$  & $0.8$ & $10$ & $0.05(1+z)$ \\
PanSTARRS & $0.75$ & $0.75$ & $5$ & $0.06(1+z)$ \\
DUNE & $0.5$ & $0.9$ & $35$ & $0.03(1+z)$ \\
\hline
\end{tabular}
\caption{Experimental specifications for the three weak lensing surveys of our analysis.}
\label{survey}
\end{center}
\end{table}

\subsection{Results}

To compute the convergence power spectrum of weak lensing we use the
code CAMB with the option 
HALOFIT \cite{Smith:2002dz} to obtain the non linear matter power
spectrum under the halo model \cite{Cooray}. Then we calculate $P_{ij}$ 
and its derivatives with respect to cosmological parameters to
construct the Fisher matrix. 
We consider the redshift range $0\leq z\leq 3$ and divide it in 5
redshift bins of equal size.

For the CMB Fisher matrix we use a modified version of CAMB, to allow
variation in $c^2_{\vis}$. 
The derivatives that appear in (\ref{fishercmb}) and (\ref{fisherwl})
are calculated from the target model. 
We use a set of nine cosmological parameters whose target values are:
$\Omega_bh^2=0.0223$, 
$\Omega_ch^2=0.114$, $\Omega_{\Lambda}=0.7$, $n_s=1.0$, $\tau=0.084$,
$w=-1$, 
$A_s=2.4\cdot10^{-9}$, $N_{\rm eff}=3.04$ and $c^2_{\vis}=1/3$. 
We assume a flat universe imposing the flatness condition:
\begin{equation}\label{flat}
h=\sqrt{(\Omega_ch^2+\Omega_bh^2)/(1-\Omega_{\Lambda})},
\end{equation}

In table \ref{sigmas} we report the $1\sigma$ uncertainties for the
parameters of our model from Fisher analysis. These results shows that
Planck alone will achieve better 
constraints on the cosmological parameters compared to weak lensing
surveys, except for the dark energy parameters $\Omega_{\Lambda}$ 
and $w$ which are very sensitive to the tomograhic information coming
from weak lensing. 
When combining CMB experiments with weak lensing surveys, the
improvement in 
$1\sigma$ uncertainties is of more than one order of magnitude on
$\Omega_{\Lambda}$ 
and $w$ and of a factor $1.5$-$2$ on the others parameters. This holds
even for 
the viscosity parameter $c^2_{\vis}$ (although this parameter has no
effect on the 
matter power spectrum and hence on the convergence power spectrum of
weak lensing) 
because of the degeneracies between the parameters of the model, such
as the degeneracies $n_s-c^2_{\vis}$ 
and $N_{\rm eff}-c^2_{\vis}$. 

\begin{figure}[htb]
\includegraphics[width=\linewidth]{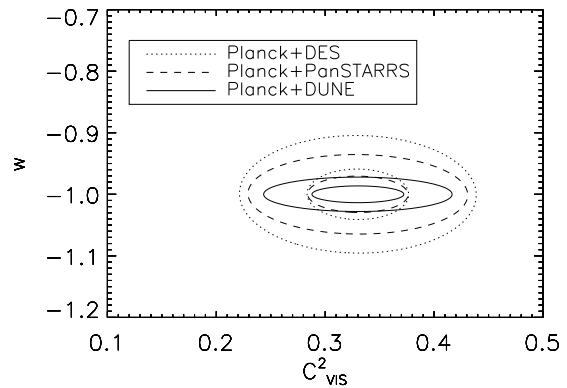}
\caption{Constraints on $c^2_{\vis}$ and $w$ from Planck combined with the three weak lensing surveys.}\label{wcvis}
\end{figure}

The combination of CMB experiments with weak
lensing  surveys can put strong constraints on $w$, though CMB alone can
contrain $w$ only weakly.  Moreover Figure \ref{wcvis} shows that there is not any relevant
degeneracy  between $w$ and $c^2_{\vis}$ and a measure of dark energy equation of
state  implies nothing for possible anisotropies in the neutrino background.

\begin{figure}[htbp]
\includegraphics[width=\linewidth]{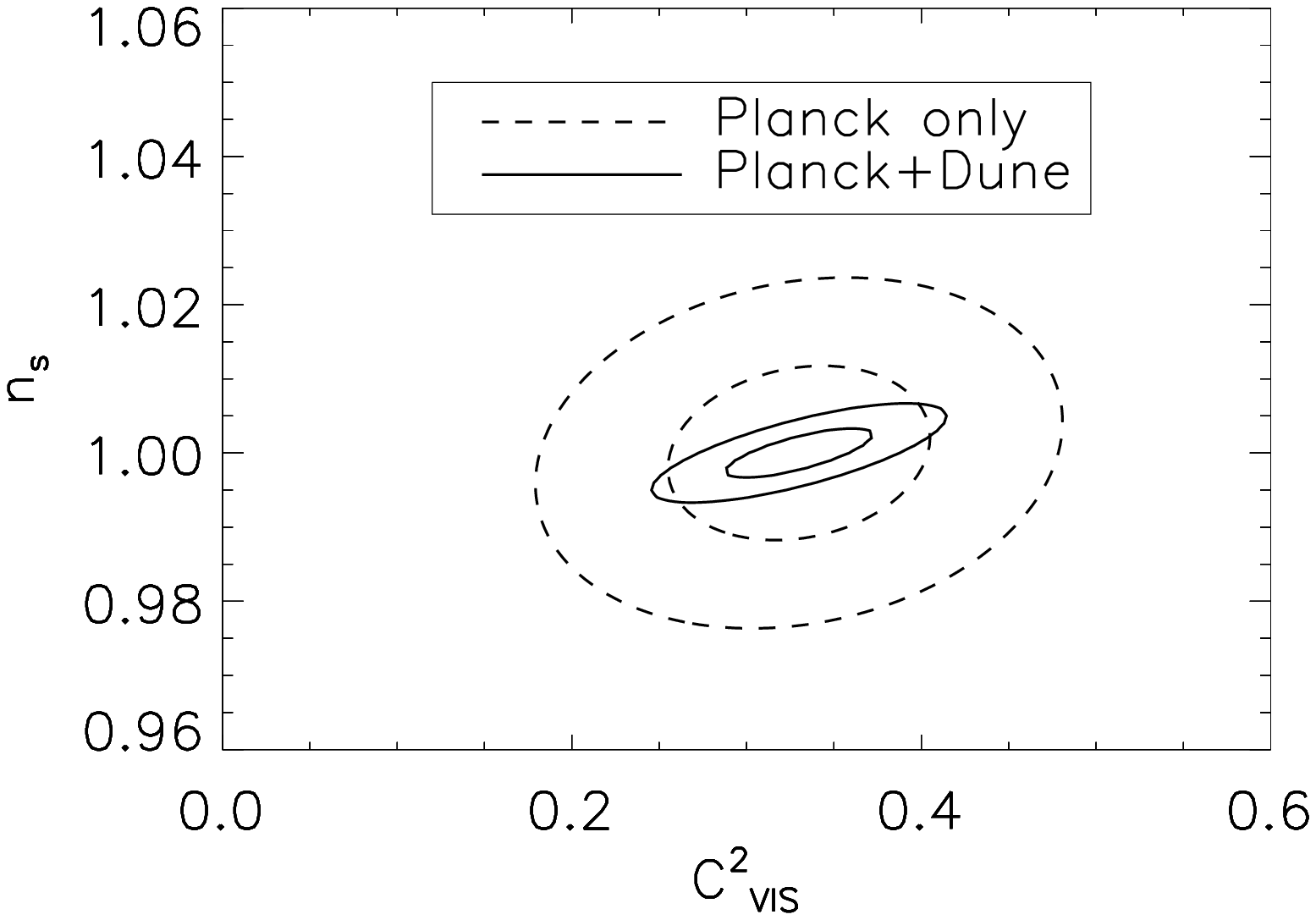}
\includegraphics[width=\linewidth]{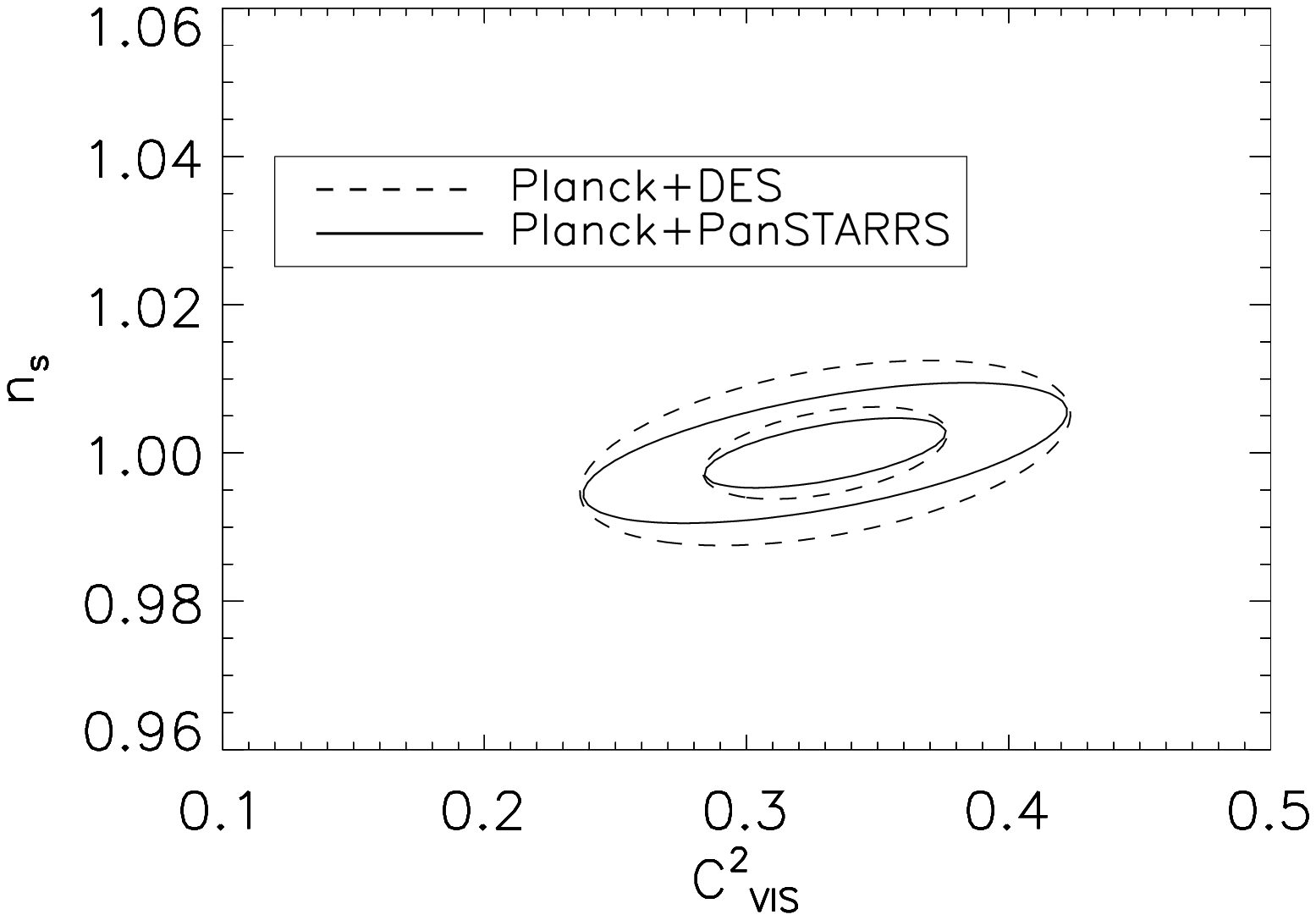}
\caption{Constraints for $c^2_{\vis}$ and $n_s$ from Planck alone and
  from Planck combined with DUNE experiment (top) and from Planck
  combined 
with others redshift surveys (bottom).}\label{ncvis}
\end{figure}

\begin{figure}[htb]
\includegraphics[width=\linewidth]{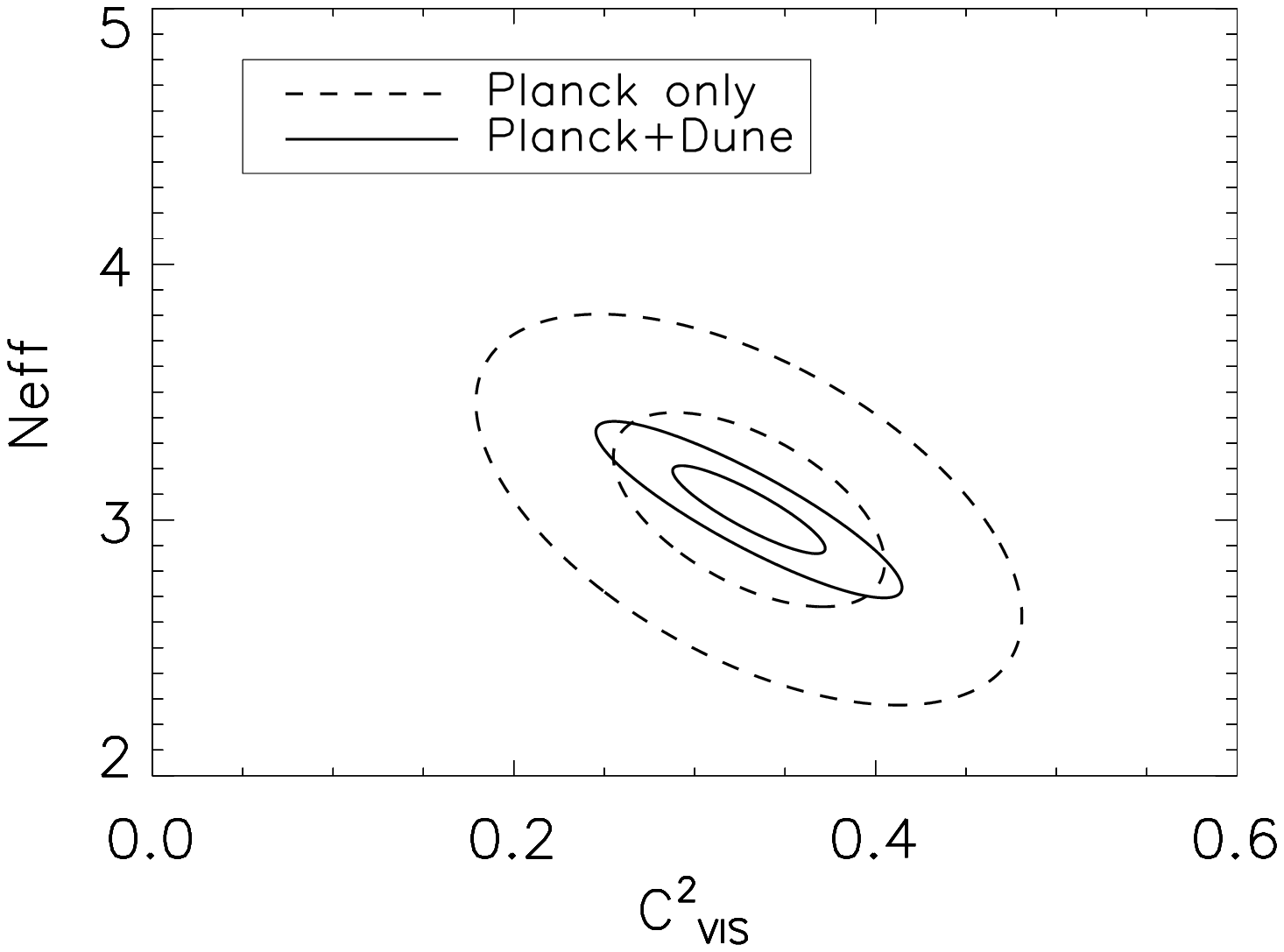}
\includegraphics[width=\linewidth]{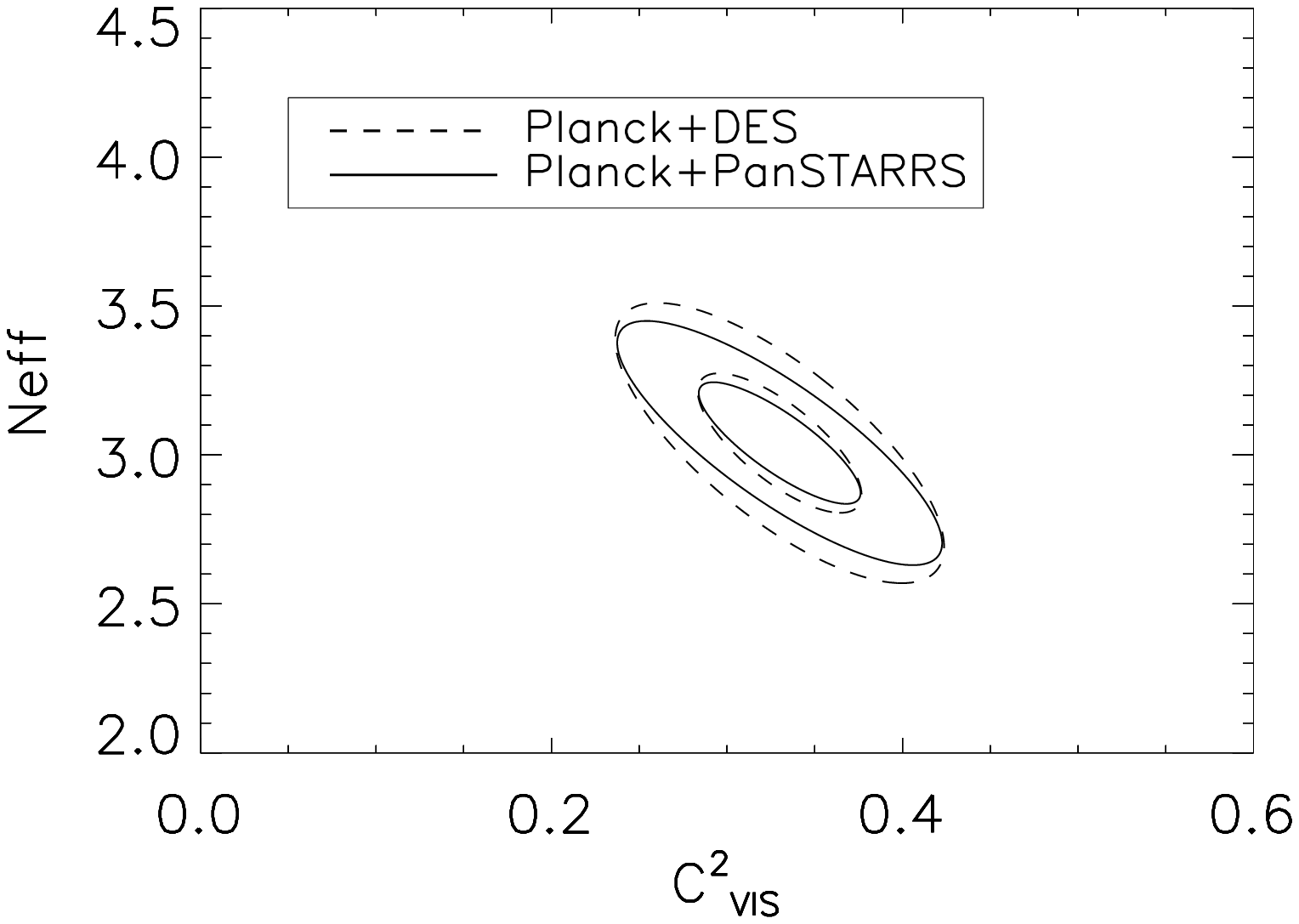}
\caption{Same as Figure \ref{ncvis} but for $c^2_{\vis}$ and $N_{\rm eff}$}\label{neffcvis}
\end{figure}

\begin{table}[ht]\footnotesize
\begin{center}
\begin{tabular}{lcccc}
\hline
\hline
&&&&\\
             & Planck      & DUNE              & PanSTARRS       & DES \\
\hline
&&&&\\
$\sigma(\Omega_bh^2)$ & $0.00044$   & $0.0059$          & $0.018$        & $0.026$\\
$\sigma(\Omega_ch^2)$ & $0.0052$    & $0.020$           & $0.057$         & $0.086$\\
$\sigma(\Omega_{\Lambda})$ & $0.12$ & $0.0051$          & $0.011$         & $0.018$\\
$\sigma(n_s)$ & $0.012$             & $0.024$           & $0.071$         & $0.10$\\
$\sigma(\tau)$ & $0.0061$           & $--$              & $--$            & $--$\\
$\sigma(w)$ &$0.39$                & $0.036$           & $0.096$          & $0.15$\\
$\sigma(\ln A_s)$ & $0.030$        & $0.088$           & $0.28$          & $0.39$\\
$\sigma(N_{eff})$& $0.38$          & $0.83$             & $2.7$          & $3.8$ \\
$\sigma(c^2_{vis})$ &$0.075$         & $--$              & $--$            & $--$\\
&&&&\\
&&&Planck+&\\
\hline
&&&&\\
&& DUNE & PanSTARRS & DES\\
\hline
&&&&\\
$\sigma(\Omega_bh^2)$ && $0.00020$     & $0.00022$           & $0.00025$     \\
$\sigma(\Omega_ch^2 )$&& $0.0032$      & $0.0036$            & $0.0039$      \\
$\sigma(\Omega_{\Lambda})$ && $0.0027$ & $0.0073$            & $0.011$     \\
$\sigma(n_s)$ && $0.0037$              & $0.0051$            & $0.0069$     \\
$\sigma(\tau)$ && $0.0054$             & $0.0057$            & $0.0058$     \\
$\sigma(w)$ && $0.015$                 & $0.031$             & $0.043$      \\
$\sigma(\ln A_s)$ && $0.012$           & $0.029$             & $0.016$\\
$\sigma(N_{eff})$ && $0.18$            & $0.21$              & $0.24$            \\
$\sigma(c^2_{vis})$ &&$0.043$           & $0.046$             & $0.047$           \\
\end{tabular} 
\caption{$1\sigma$ errors on the cosmological parameters of our model from Planck and weak lensing surveys alone (top) and from Planck combined with each survey (bottom).}\label{sigmas}
\end{center}
\end{table}
In Figure \ref{ncvis}, Figure \ref{neffcvis} and Figure \ref{wcvis} we
show the joint constraints on $c^2_{\vis}$  
with $n_s$, $N_{\rm eff}$ and $w$, from Planck alone and from Planck combined with the three weak lensing survey.

To obtain the confidence contours for any couple of parameters, we
take the elements of the inverse Fisher matrix
$(F^{-1})_{\alpha\beta}$ that corresponds to those parameters. This
gives the correlation matrix of that couple of parameters and the
eigenvalues and eigenvectors of this correlation matrix give the
orientation and the size of the ellipsoid of confidence. This
procedure is equivalent to marginalize on the remaining parameters
(see also \cite{Sapone:2007jn}). 

In performing the Fisher analysis we are assuming that the likelihood
is Gaussian and that the maximum of the likelihood is in the target
model. The likelihood for the parameter $c^{2}_{\vis}$ could not be
exactly Gaussian,  at least for $c^2_{\vis}>1/3$. However expanding the likelihood in
Taylor series around  the target model we can see that the likelihood function is
approximately Gaussian for $c^2_{\vis} \le 1/3$,
(i.e. it is dominated from quadratic terms), even if it could be not
Gaussian globally. In fact usually $L$ drops 
rapidly so that we can neglect the non-Gaussianity of the likelihhood
\cite{Tegmark:1996bz} and the constraints 
from Fisher analysis can be considered a good approximation. 

The most considerable result of this analysis is that it is possible to detect
neutrino background anisotropies  even from Planck alone with
$0.33>c^2_{\vis}>0$.  The combination of Planck experiment with weak lensing surveys can
decrease the statistical 
uncertainty on $c^2_{\vis}$ about of a $35\%$ (for Planck+Dune), while we find
the combinations Planck+DES and  Planck+PanSTARRS give similar constraints.

\section{Conclusions}

In this paper we have re-analyzed the status of neutrino anisotropies
in light of the new WMAP five-year data. We found that while
the WMAP 5-year data alone are unable to provide significant evidence for
those anisotropies, combination of CMB data with current large-scale structure cosmological data yields
a detection at more than $2 \sigma$ confidence level.
Future cosmological data are certaily needed to confirm this result.

In this respect we have performed a forecast for future CMB and
weak lensing missions. We have found that a combination of DUNE and
Planck will be able to yield a measurement of the $c_{\vis}$
parameter at $\sim 10 \%$ level. Moreover, we studied the possible degeneracies with other cosmological
parameters. Including uncertainties on $c_{\vis}$ will double the
error bars on $N_{\rm eff}$ while the constraints on the equation
of state $w$ will remain pratically unaffected.
While the standard model predicts $c_{\vis}^2=1/3$ several cosmological
scenarios, from interacting neutrinos to early dark energy, can
be considered that could bring a deviation in the measured value.
A better detection of anisotropies in the neutrino background could 
therefore provide an useful test for those models and possibly
indicate the presence of new physics.

\textit{Acknowledgements}
It is a pleasure to thank Alan Heavens, Roberto Trotta and Licia Verde 
for helpful discussions.
This work was supported by NSF CAREER AST-0645427 at UC Irvine. 
This research has been supported by ASI contract I/016/07/0 "COFIS".

\end{document}